# EXPERIMENTAL STYLOLITES IN QUARTZ AND MODELED APPLICATION TO NATURAL STRUCTURES

**J.P. Gratier, L. Muquet, R. Hassani, F. Renard**

J. P. Gratier: LGIT-CNRS-Observatoire, Université Joseph Fourier, BP 53, 38041 Grenoble, France

Fax: +33 476 82 81 01

gratier@lgit.obs.ujf-grenoble.fr

# Abstract

Experimental stylolites have been observed at stressed contacts between quartz grains loaded for a period of several months in presence of aqueous silica solution, at 350°C under 50 MPa of differential stress. Stereoscopic analysis of pairs of SEM images, processed in the same way as earth-surface elevation data gives the stylolites topography. Coupled with observations of closed interactions between dissolution pits and stylolitic peaks, these data illuminate the mechanism of stylolite formation. The complex geometry of stylolite surfaces is imposed by the interplay between the development of dissolution peaks in favored locations (fast dissolution pits) and the mechanical properties of the solid-fluid-solid interfaces.

Simple mechanical modeling expresses the crucial competition that could rule the development of stylolites: (i) a stress related process (modeled as the stiffness of springs (N/m$^3$) activates the heterogeneous dissolution rates of the solid interface that promotes the deflection. In parallel, (ii) the strength of the solid interface, modeled as the stiffness of a membrane (N/m) and equivalent to a surface tension) limits the deflection and is opposed to its development. The modeling produces stylolitic surfaces with characteristic geometries that vary from conical to columnar shaped stylolites when both the effect of dissolution-rate heterogeneity and the strength properties of the rock are included.

# Introduction

Stylolites are well known as natural markers of pressure solution. Following the interpretation of (Stockdale, 1922), that related stylolite development to stress-driven dissolution, there has been general agreement about such an explanation (Dewers and Ortoleva, 1990b), (Dunnington, 1954), (Heald, 1955), (Park and Schot, 1968), (Arthaud and Mattauer, 1969), (Carrio-Schaffhauser et al., 1990), (Dewers and Ortoleva, 1990a). Nonetheless, some authors also suggested that cataclastic deformation could contribute to the specific stylolitic shape (Deelman, 1976), (Milliken, 1994). Based on the observation of roughly regular spacing, the development of stylolitic layering surfaces has been related to self-organized processes (Merino et al., 1983), (Dewers and Ortoleva, 1990a), with such processes arising from instability in chemically compacting rocks and leading to the formation of spatial heterogeneity in rock. However, this type of model has been questionned by other authors who consider either that the spacing of stylolites is indistinguishable from random arrangement (Railsback and Andrew, 1995) or that initial conditions (e.g., fracture patterns) could impose a rough regular spacing (Railsback, 1998). The intriguing stylolitic tooth-like shape itself has been interpreted as a self-organized process linked to the interplay between stress state evolution and progressive change of porosity of the rocks (Gratier, 1987). Due to the well-recognized role of stylolites in compaction and creep processes (Bathurst, 1971), it is important to decipher the effect of the various parameters in their development. Much of the previous studies have been done on natural stylolitic structures. In such cases, information on crucial parameters, mainly those concerning initial state, mechanical behavior, stress-state and dissolution-rate are missing. This is why trying to obtain experimental stylolites by controlling primary parameters in the lab could help understanding their development. However, very slow kinetics of pressure solution processes (Rutter, 1976) make them very difficult to reproduce. We present here the results of a successful attempt to obtain experimental stylolite surfaces. Such experimental stylolites, at the grain scale, are developed to the extent that they can be studied as carefully as natural stylolites surfaces. The geometric analysis of such experimental stylolite surfaces leads us to propose and to test a conceptual model for stylolite development.

# Experiments

Transparent euhedral quartz crystals collected from tension gashes in Oisans massifs (French Alps), were grounded and a 100-125 microns fraction selected. Sand layers of those fine quartz grains were sandwiched between cylindrical samples of quartzite, flooded with a silica solution, then loaded during several weeks under a vertical uniaxial stress ($\sigma_l$) of 200 MPa. Experiments were run in pressure vessels at temperature of 350°C and fluid pressure ($P_f$) of 150 MPa. Thin sections of the deformed sand layers were analyzed in order to study the effect of various parameters (stress, nature of the solution) on the kinetics of pressure solution, (Gratier and Guiguet, 1986). Experimental pressure solution strain-rates appeared to be limited by diffusion-rates along stressed grain boundaries. This kinetic aspect is not developed any further here. In the present paper we focus on scanning electron microscope (SEM) studies of the stressed grain contact that reveal micron-scale stylolites along stressed grain to grain contacts. The stylolite samples were obtained from two types of experiments differencing in the nature of the solution and the duration of the two experiments, and being respectively: pure water for 51 days and aqueous NaOH 0.1 mole/l solution for 43 days.

A general view of quartz sand layers after deformation is shown in Fig. 1a. SEM photographs were taken directly on such samples. At this millimeter scale, small holes are visible in the initially polished quartzite support of the sand layer. When observed in thin sections, these holes were clearly related to localized dissolution along grain boundaries, by sand grain indenters (Gratier and Guiguet, 1986). In Fig. 1b, a stylolite surface appears well developed on the top of the grain, with peaks that are parallel to the maximum compressive stress ($\sigma_1$). A slickolite (dissolution surface oblique to the direction of displacement) is developed on the same grain in lateral continuity to the stylolite surface with its associated spikes also parallel to $\sigma_1$. Such patterns are observed only at grain contacts, the free of stress surfaces being flat and smooth. A more complex evolution of the grain aggregate appears in Fig. 1c where a stylolitic surface abruptly evolves into slikensline structure. This evolution must have happened due to the slight rotation of one of the grains after the development of the stylolite. Most often, the free of stress surfaces of the grains which are more or less parallel to $\sigma_1$ (grain boundaries or vertical fractures) show no evidence of stylolites (Fig. 1b, 1c). Such near vertical surfaces, looking into pore space, often open during the sand compaction and are filled with small euhedral quartz crystals (Gratier and Guiguet, 1986). However, in some very rare cases, stylolites with horizontal peaks are also developed on the vertical side of some grains, indicating local compressive stress oblique to the maximum one. As the already mentioned stylolite-to-slikensline transition, such heterogeneous behavior may be related to a complex rearrangement of initially irregular grains during the progressive deformation. All over the samples, it can be seen that stylolites develop both on top and bottom of each pair of stressed grains (Fig. 1d and 1 e), in an apparent symmetrical way, and without any visible mismatch along the seam. However, enlarging the view of the stylolitic surfaces reveals networks of dissolution pits that cross cut the whole dissolution surfaces (Fig. 1e). These pits may be found all along the stylolitic surfaces. However, there is a systematic location of the pits, which appear at the bottom of each conical-shaped indented stylolitic structure, which consequently looks like a funnel-shaped structure. Due to the fit of two opposite surfaces, the pits of the lower grain stylolitic surface are located just in front of the stylolitic peaks of the upper grain and vice versa. Detailed photographs of the pits (Fig. 1f) reveal their negative euhedral crystal shape, which ressemble equilibrated fluid inclusions in quartz.

We have performed stereoscopic analysis of pairs of SEM photographs, processed in the same way as topographic surfaces, which yielded topographic data sets for two of these stylolitic surfaces. One of these surfaces is represented on Fig. 2b and 2c, in map and perspective view respectively. Fourier analysis of the surfaces (Fig. 2d) indicates that the dominant spatial wavelength on the two surfaces ranges between 2.2 and 3.1 microns. Mean slopes of the peaks and the depressed areas are equal on the two surfaces (around 36°) and there is no slope steeper than 65° (Fig. 2e).

## Discussion of the experiments

At the grain scale (Fig. 1b), the parallelism between slickolite spike and stylolite peak implies that the relative displacement vectors of the dissolution surface between two stressed grains are vertical, and parallel to $\sigma_1$. However, at the micron scale (Figs. 1d-e) the mean slope of the stylolitic surface is oblique to this displacement. This means that stress driven dissolution occurred, at least partially, on surfaces that were oblique to the main loading uniaxial stress. This observation is not specific to experimental stylolites. Observations of natural stylolites show a large geometrical variability (Park and Schot, 1968). Columnar stylolites have the simplest geometry, with column sides parallel to the displacement and column bases perpendicular to $\sigma_1$, (Fig. 3a). However, natural conical-shaped stylolites are also common (Fig. 3b). Despite four orders in magnitude difference in size, such a natural stylolitic geometry is similar to the experimental one (respectively Fig. 2c and 3b). It has been suggested that columnar stylolites result from coaxial stress history whereas conical shaped stylolites are thought to result from a non-coaxial stress history. Our experiment show that conical stylolites can form from a coaxial history, therefore a non-coaxial history is not necessarily required for their formation. Observations at the maximum enlargement (Fig. 1f) also give information on the geometry of the grain-to-grain fluid interface: no evidence of channel and island structure may be found at this scale. The trapped-fluid interface may be best qualified as "pitted and waffled water-film".

The striking feature of such experimental stylolites is the systematic link between stylolitic peaks and dissolution pits. Pits may be seen all over the surface, however they are systematically located in front of each peak (see Fig. 2a), with a polyhedral shape related to the crystalline structure (Fig. 1f). Such etch pits are commonly observed on mineral surfaces which have undergone aqueous dissolution. Such features have been shown to develop at the intersection of dislocations with the mineral surface (Frank, 1951), (Blum et al., 1990), (Brantley et al., 1986). Enhanced dissolution-rate is presumably linked to high strain energy stored around the dislocations or to localized attack by the solvent on the geometric step on the crystal surface (Bosworth, 1981), (Tada and Siever, 1986). Low dislocation density commonly appears during the crystal growth, whereas high density may develop during plastic deformation. For natural quartz, these two processes typically lead to dislocation densities ranging respectively from $10^5$ to $10^{10}$ dislocations cm$^{-2}$, (Blum et al., 1990). When assuming a plausible link between pits and dislocations, a rough analysis of etch pits density in Fig. 2a indicates a density of about $10^8$ dislocations cm$^{-2}$. Lines of high pits density are seen in the photograph on Fig. 2a, and could reveal planar defects in the crystal more or less related to the crystalline structure. The systematic face-to-face coincidence of stylolitic peaks and dislocation pits accounts for several explanations. Dislocation pits may predate stylolitic development and their distribution could play a role in the localization of the peak. In this case, heterogeneities in the crystal would control the development of the stylolites. Dislocation pits may also develop after stylolitic amplification due to an indenter effect of the peaks. However, in such cases, they could also play the role of favored dissolution sites as soon as they develop. Moreover, some pits must predate the stylolitic development since pits are found all along the slope of the stylolites peaks (Fig. 2a). So for simplicity, we will consider that when two stressed solids are pressed in contact, with their solution trapped in the interface, dissolution-rates are heterogeneous and some pits develop rapidly that can initiate a slight deflection of the interface. Such deflections concentrate the stress and may amplified the development of the pits in front of each growing peaks. This is a stress-induced instability, as developed by Kassner et al. 2001.

## Principle of the modeling

We propose a simple modeling approach, which is able to generate stylolitic surfaces with characteristic natural geometry ranging from conical-shaped to columnar-shaped stylolites (Fig. 3), as well as experimental geometry (Fig. 2). When considering the geometry of the surfaces of dissolution, there is a difference between conical and columnar stylolites : dissolution interfaces of columnar stylolites are disrupted by cinematic discontinuities that are parallel to the displacement (analogous to transfer faults), whereas conical stylolites show continuous (deflected) surface of dissolution (Fig. 4b). However, transfer faults may also be associated with conical stylolites when their size is larger than the peaks size (such as the cinematic discontinuity seen in Fig. 1b, and schematically drawn in Fig. 4c).

The principle of our model is based on the experimental observation that dissolution pits are always located opposite to the stylolites peaks. Let consider a fluid phase trapped between two solids under stress (Fig 4d). Let also consider that stress induced dissolution-rates are heterogeneously distributed along the two dissolution surfaces. For example in experiments, dissolution rates near dislocations pits are higher than dissolution rates away from the dislocations pits presumably because of local stress concentration. In natural deformation, favored zone of dissolution may derive from initially high porous area that determined the dissolution location, as in the self-organized model of stylolitic layering (Merino et al., 1983), (Dewers and Ortoleva, 1990b),. It may also derive from zones of high fracture density that are known to significantly enhance dissolution (Railsback, 1998), (Gratier et al., 1999).

At all scales, fastest dissolution sites along one of the surface generate a small deflection of the opposite surface due to the main compressive stress perpendicular to the solid/fluid/solid interfaces. The small deflection generates stress concentration in the solid in a direction opposite to the tip of the deflection (Jaeger and Cook, 1969). This increases the dissolution-rate in the vicinity of the already favored zone of dissolution and this leads to the progressive growth of the stylolite peaks as a self-organized process. If both solid surfaces dissolve in a symmetrical way the shape of the stylolitic surface result from a complex interaction between the two solids and the dissolution-rates distribution.

A schematic view showing the evolution of stylolites based on observations of experiments is given in Fig. 4 d, e, f. Points A, B, C are outcrops of dissolution pits along the two grain interfaces: dislocations A and B being in the lower grain and C in the upper grain. The dissolution trajectories are assumed to remain parallel to the general $\sigma_1$ direction. The various dissolution rates associated with different dislocation pits and the random distribution of these dislocation pits lead to a complex evolution of the stylolite surface. For example (Fig. 4 f-top), dislocation pit B (lower grain), with a relatively slow dissolution-rate versus those of dislocation C (upper grain) is moved up to the top of the stylolite peak associated with dislocation C. Such a geometry is well recognized in experiments (Fig. 1d & 2a). However, if site B and C are both very active (fast dissolution, as in Fig. 4f bottom) a cinematic discontinuity may develop especially if the strength of the nearby solid is weak. This is seen in the experiment (Fig. 1b) were the slickolite surface ends up along transfer fault (model 4c). This disposition is also common in natural stylolites when microfractures cut the rocks (see natural examples, Fig. 3c).

Consequently, the modeling must include the strength of the solid interface around the favored zones of dissolution. We used a simplified approach by considering the mechanical equilibrium of a thin elastic membrane submitted to opposite vertical forces. The membrane is deformed by positive or negative incremental displacements of localized patches representing the fast displacement dissolution areas in both the upper and the lower grains respectively. The displacement rates modeled the heterogeneous dissolution rates distribution along the solid interface that promotes the deflection. In parallel, the strength of the membrane limits the deflection and is opposed to its development. The principle of the model may be matched with the principle of the development the Grinfeld instability that describes the behavior of solid surface of which material transport is possible and that are under compressive stress parallel to the surface. In this case Grinnfeld & Hazzledine (1996) and Kassner et al. (2001) showed that when accidental corrugation develop at the solid/fluid surface two processes are competing to increase or anneal the deflection: elastic energy stored in the solid tends to develop the deflection of the surface whereas surface energy limits the development of a rough surfaces.

The modeling algorithm is derived as follows:

1 Fast dissolution patches of fixed diameter "d" (Fig. 4c), are randomly distributed through the entire surface.

2 Half of these patches are associated with upward displacement (upward forces), the other half being associated with downward displacement (downward forces).

3 At each incremental step of the loading process, only randomly tossed patches are displaced in order to simulate the heterogeneity of the dissolution-rates distribution. This is similar to the principle of the "game of life" models (Gardtner, 1970), in particular when used to model self-organized systems (Bak and Tang, 1989).

4 The elastic properties of the solid are kept constant throughout the whole process.

At each increment, the thin elastic membrane is loaded by vertical forces and submitted to buoyant restoring forces on its two faces by elastic springs. The vertical displacement field $u(x,y)$ of the membrane is then governed by the differential equation:

$$F = \tau (d^2u/dx^2 + d^2u/dy^2) + k\, u$$

Where F is the vertical density of force (N/m$^2$); $\tau$ is the stiffness of the membrane (N/m) and is equivalent to a surface tension; and k is the stiffness of the springs (N/m$^3$) as a force per unit of volume. The stiffness of the springs k represents the characteristics of the dissolution process (ratio F / u, if $\tau$ is negligible), the stiffness of the membrane $\tau$ represents the stiffness of the rock along the dissolution interface. The spatial load distribution F(x,y) represents the random geometry of the fast dissolution patches. The ratio k / $\tau$ and the spatial distribution of the patches F(x,y) allow reproduction of the various stylolite geometries.

## Discussion of the models

Results of the modeling are presented in Fig. 5a and 5b. Two types of structures are obtained which are representative of two types of stylolite geometry. Conical stylolites developped in several increments, for example from 400 initial small patches (half with positive forces, half with negative ones) as shown in Fig. 5a. Perspective views show the progressive development of such stylolites surfaces. Specific characteristic of experimental stylolites can be seen. The largest stylolite peaks and conical depressed areas develop in front of the most active dissolution pits (positive and negative respectively). Secondary peaks and depressed areas of smaller size are seen all along the surface corresponding to less active dissolution sites. A complex organization emerges from such a simple modeling that will be analyzed more carefully in the future. With such a simplified mechanical modeling, the vertical displacement is representative of the ratio between force and stiffness. Consequently, investigating the detail geometric characteristics of the theoretical surface is not a simple matter and was beyond the scope of this paper. With such simplified modeling we simply tried to reproduce the main characteristics of the stylolites with the minimum number of parameters namely two main parameters: heterogeneous dissolution-rate distribution (spatial load distribution) and simplified mechanical properties of the solid (membrane stiffness). Results of the modeling (Fig. 5) show the effect of these two parameters on the geometry of stylolites. Columnar stylolites (Fig. 5b) are obtained both when using a smaller number and a larger size of the patches of fast dissolution and when using a much lower stiffness of the elastic membrane than for conical stylolites (Fig. 5a). This clear difference tells us information about the dissolution sites heterogeneity (size and distribution) and the mechanical behavior of the rocks during stylolites development. According to our model, columnar stylolites indicate a much lower stiffness of the grain interface than conical stylolites. This low stiffness may be associated either with poorly lithified sediment at the beginning of the diagenetic process, or it may be associated with tension fracture parallel to the maximum compressive stress (Fig. 3c) that has no shear strength parallel to the potential transfer fault. The processes that impose the fast dissolution patches probably vary with the size of the stylolites. It could be dislocation related dissolution pits for micron size (Fig. 1). It could be also some zones of high porosity or high fracture density at millimeter size (Fig. 3). More complexity is awaited if one considers a solid developing conical stylolites (numerous localized dissolution patched and high stiffness of the solid interface) when the solid include preexisting fractures that may evolve to cinematic discontinuity.

## Conclusion

Experimental stylolites have been observed at stressed contacts between quartz grains loaded for a period of several months in presence of aqueous silica solution, at 350°C under 50 MPa of differential stress.

SEM images and the derived stylolites topography reveal that the complex geometry of stylolite surfaces is imposed by the interplay between the development of dissolution peaks in favored locations (dissolution pits) and the mechanical properties of the solid-fluid-solid interfaces.

Simple mechanical modeling produces stylolitic surfaces with characteristic geometries that vary from conical to columnar shaped stylolites when both the effect of dissolution-rate heterogeneity and mechanical properties of the rock are included.

Experiment and discrete modeling allow distinguishing the role of two main parameters in the development of stylolitic geometry.

1 - The initial distribution of heterogeneous dissolution-rate areas along the interface localize the first stylolite peaks and control the progressive deflection of the surface of dissolution.

2 - The mechanical property of the stressed solid imposes the slope of the dissolution surfaces around the favored zones of dissolution.

The model expresses a crucial competition that may rule the development of stylolites:

1 – A stress related process (modeled as the stiffness of springs ($N/m^3$) activates the heterogeneous displacement rates of the solid interface that promotes the deflection.

2 - The strength of the solid interface, modeled as the stiffness of a membrane ($N/m$) and equivalent to a surface tension) limits the deflection and is opposed to its development.

Interaction between this two main parameters lead to various geometrical shape of stylolites surfaces. Conical stylolites develop with high stiffness of the solid and with relatively localized patches of fast dissolution (as dissolution pits in experiments stylolites or localized spots in natural stylolites). Columnar stylolites develop with low stiffness of the solid and with relatively large patches of fast dissolution patches (for example in relation with heterogeneous compaction or fracturating process of the rock). Mixted model of conical stylolites associated with fault transfer may also develop with high stiffness solid that are cut by tensile fracture which evolve into cinematic discontinuity.

# Figures

Figure 1

Experimental stylolites developed by stress driven dissolution process in presence of various solutions: pure water (c, f) and Na OH 0.1 M (a, b, d, e); see other parameters in the text. a: General view of quartz sand layer after deformation. b: Stylolite and slickolite on a single grain with peaks and spike parallel to $\sigma_l$. c: Stylolitic surface that abruptly evolves into slickenline structure after local grain rotation. d & e: Stylolite on both bottom and top stressed grains respectively. f: detail of dissolution pits that develops in front of each stylolite peak.

Figure 2

Geometrical analysis of stylolites. 2a: Stereoscopic SEM photograph, view parallel to experimental stylolite peaks. 2b: Structure contour map from stereoscopic restitution. 2c: 3D perspective view of experimental stylolite from stereoscopic restitution. 2d: Spatial wavelength analysis of the stylolite surface by Fourier transform. 2e: Cumulative distribution of the slopes on two different experimental stylolite surfaces.

Figure 3

Natural stylolites, with their shape varying from columnar stylolites (3a) to conical styloliutes (3b). 3c: Details of the shape of some columnar peaks, the column being bounded by microfractures.

Figure 4

Schematic view of stylolites development, dissolution surfaces are red and micro-tear faults are blue. 4a : columnar stylolites (the dissolution surface is interrupted by with micro-tear faults. 4b: conical stylolites (the dissolution surface is continuous). 4c: conical stylolites associated wit transfer faults (schematic view of the Fig 1b). 4d to f : cinematic model of stylolite development as favored dissolution sites (of diameter d) determine the stylolite peaks location: two possibilities are presented with cinematic discontinuity as transfer fault (bottom) or without such a discontinuity (top).

Figure 5

Simplified mechanical modeling of stylolite process. 5a: modeled conical stylolites with 4 successive increments of loading, on the left is the random geometry of the patches of fast dissolution at each step (the red ones are associated with upward displacement, the blue ones are associated with downward displacement). On the right are perspective views of 2 steps (the first one and the fourth one). 5b: Modeled columnar stylolites, on the left are the perspective views, on the right are the distribution of the favored dissolution sites (the color chart has the same meaning as in Fig. 5a).

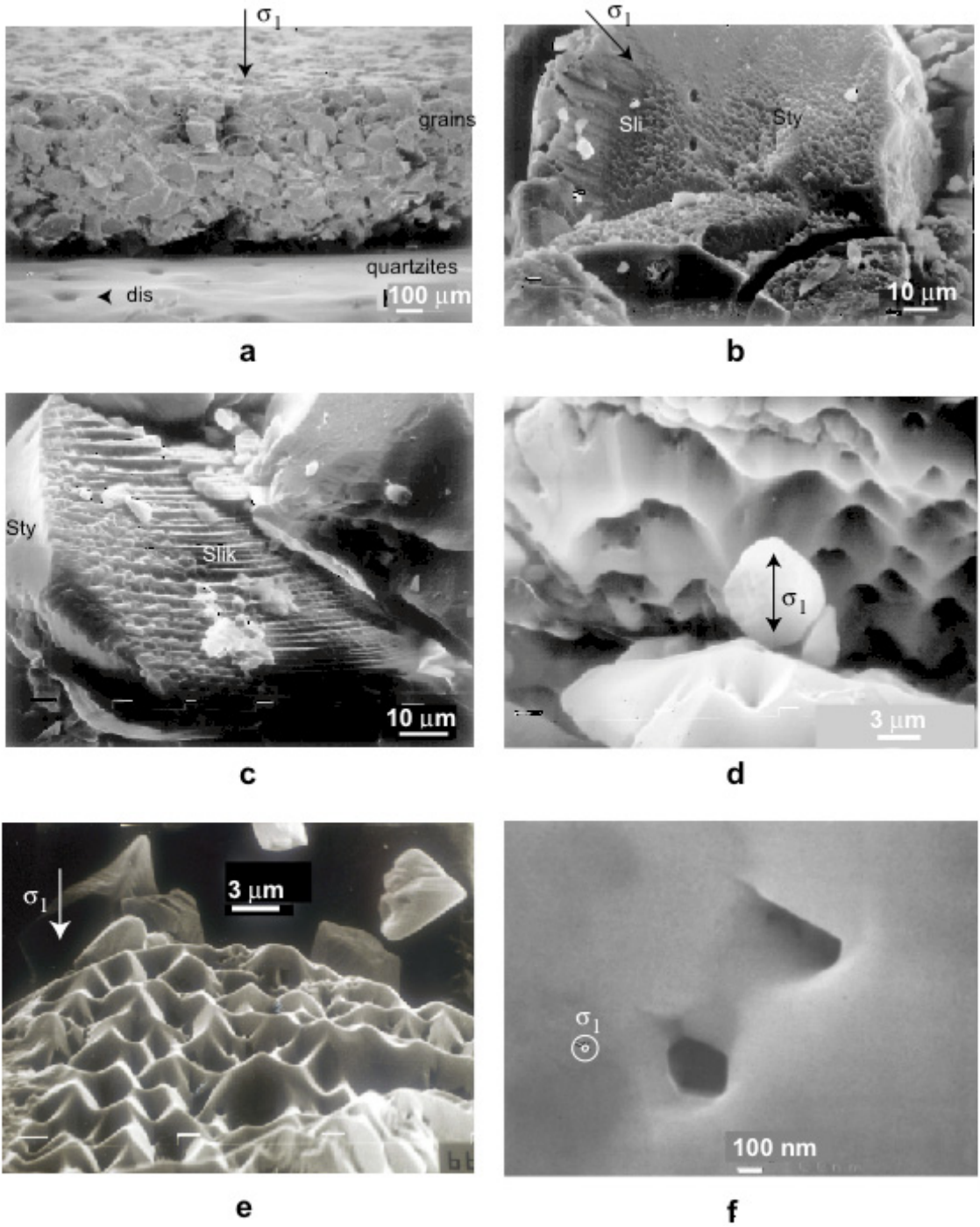

Figure 1

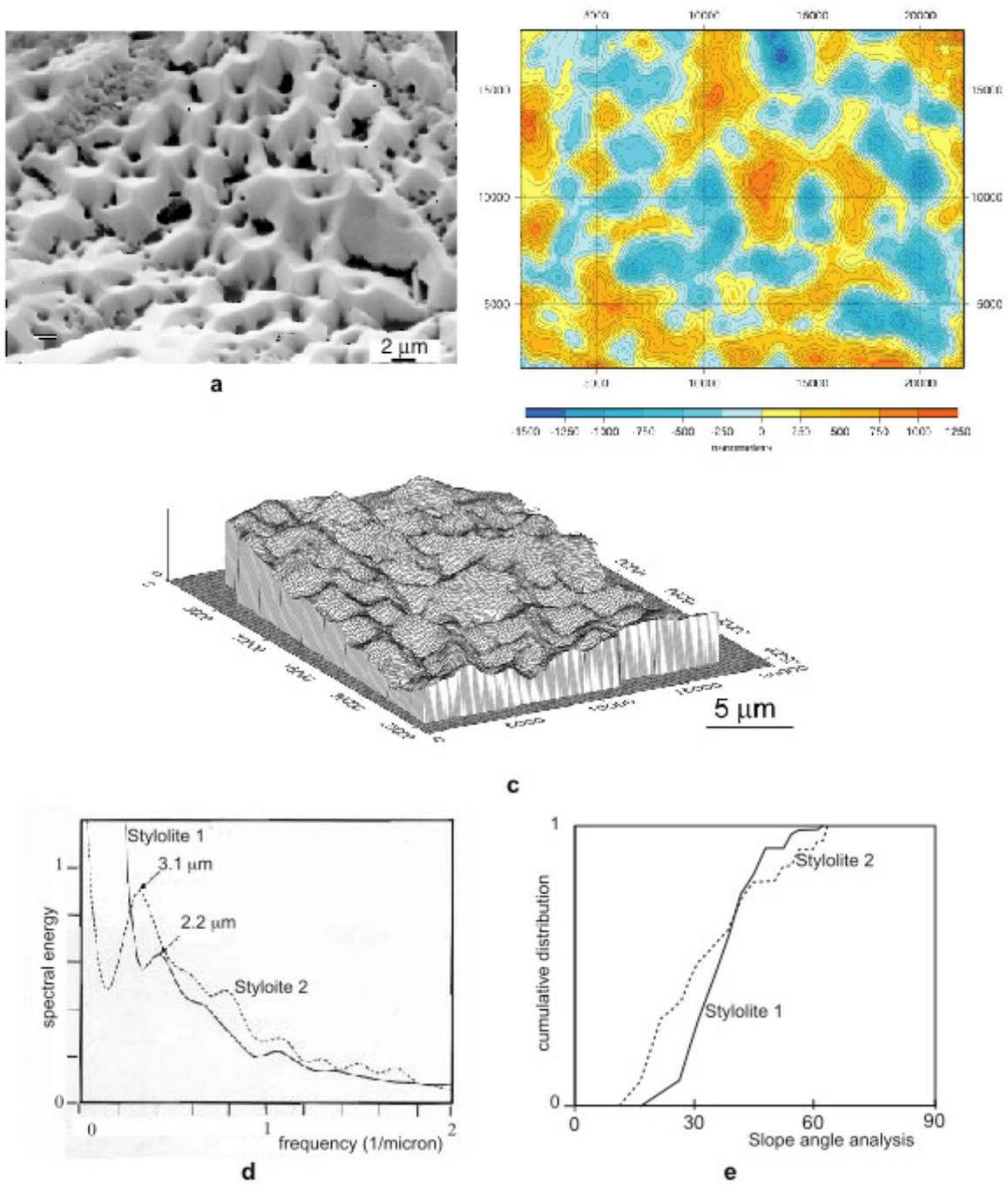

**Figure 2**

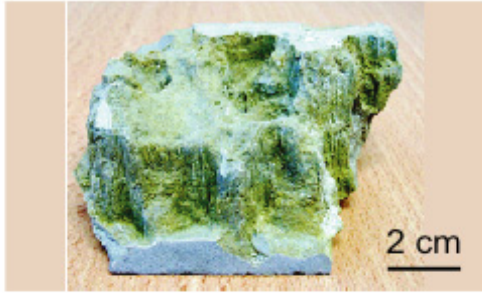
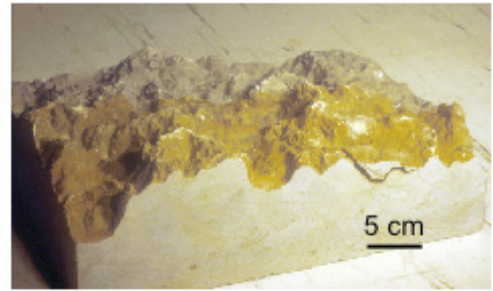

a

b

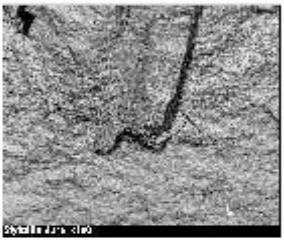
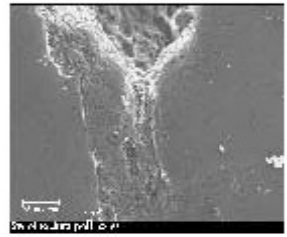
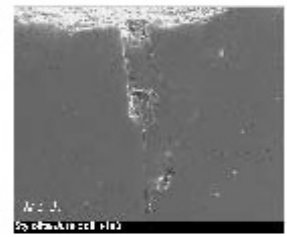

figure 3

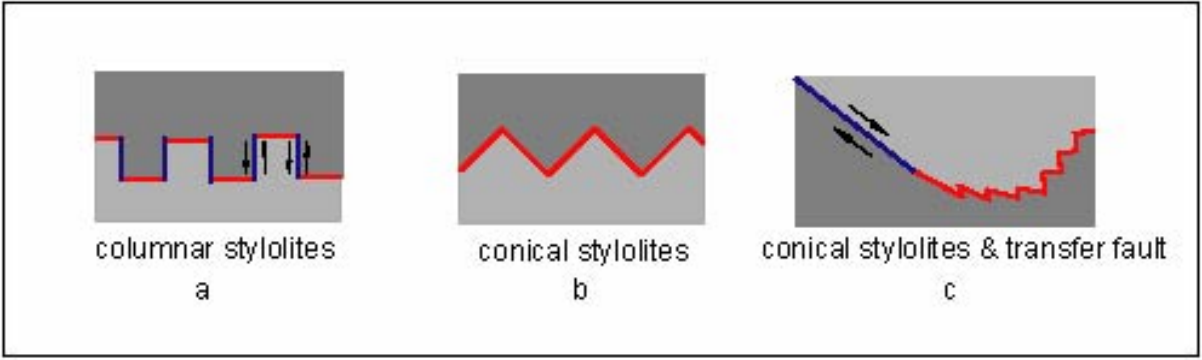

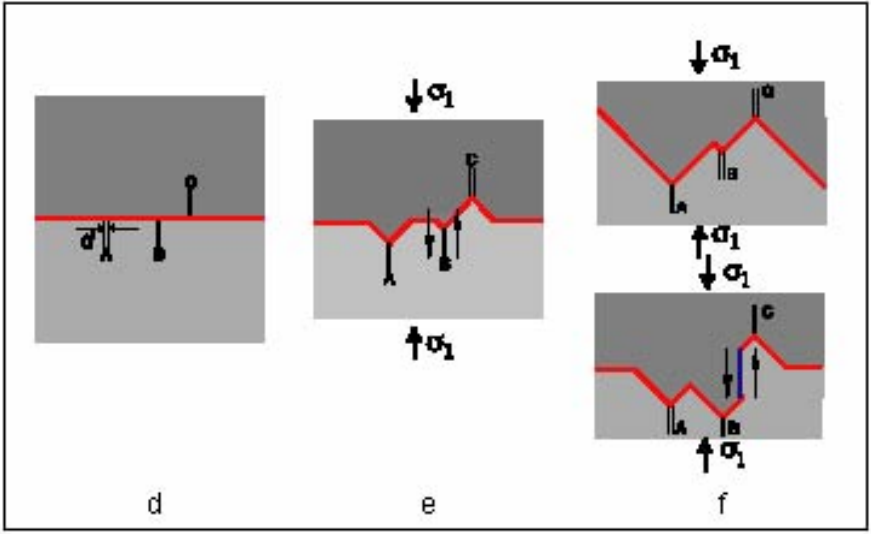

Figure 4

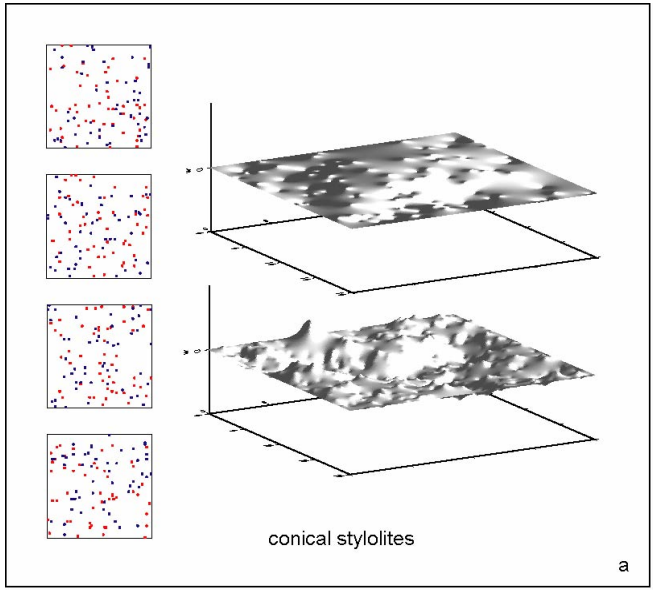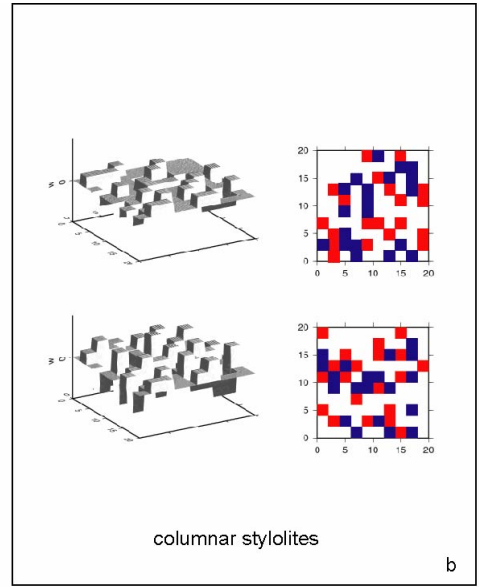

Figure 5